\documentclass{pasj00}

\begin{document}
\SetRunningHead{N. Inada et al.}{Central Structure of SDSS J1004+4112}
\Received{2008/07/24}
\Accepted{2008/07/25}

\title{Spectroscopic Confirmation of the Fifth Image of SDSS
  J1004+4112 and Implications for the $M_{\rm BH}$-$\sigma_*$
  Relation at $z=0.68$\thanks{Based on data collected at 
  Subaru Telescope, which is operated by the National Astronomical
  Observatory of Japan, and observations (GO-9774, GO-10509, and 
  GO-10716) made with the NASA/ESA 
  Hubble Space Telescope, obtained at the Space Telescope Science 
  Institute, which is operated by the Association of Universities 
  for Research in Astronomy, Inc., under NASA contract NAS 5-26555. 
  }
} 

\author{
Naohisa \textsc{Inada},\altaffilmark{1}
Masamune \textsc{Oguri},\altaffilmark{2}
Emilio E. \textsc{Falco}, \altaffilmark{3}
Tom J. \textsc{Broadhurst},\altaffilmark{4}
Eran O. \textsc{Ofek},\altaffilmark{5} \\
Christopher S. \textsc{Kochanek},\altaffilmark{6}
Keren \textsc{Sharon},\altaffilmark{4} and
Graham P. \textsc{Smith}\altaffilmark{7} 
}
\altaffiltext{1}{Cosmic Radiation Laboratory, RIKEN, 2-1 Hirosawa, 
Wako, Saitama 351-0198.}
\altaffiltext{2}{Kavli Institute for Particle Astrophysics and
Cosmology, Stanford University, Menlo Park, CA 94025, USA.}
\altaffiltext{3}{Harvard-Smithsonian Center for Astrophysics, 60 Garden 
Street, Cambridge, MA 02138, USA.}
\altaffiltext{4}{School of Physics and Astronomy, Tel Aviv University, 
Tel Aviv 69978, Israel.}
\altaffiltext{5}{Department of Astronomy, 105-24, California Institute 
of Technology, Pasadena, California 91125, USA}
\altaffiltext{6}{Department of Astronomy, The Ohio State University, 
Columbus, OH 43210, USA.}
\altaffiltext{7}{School of Physics and Astronomy, University of Birmingham, 
Edgbaston, Birmingham, B152TT, UK.}


\maketitle

\begin{abstract}
We present the results of deep spectroscopy for the central region of 
the cluster lens SDSS~J1004+4112 with the Subaru telescope. A secure
detection of an emission line of the faint blue stellar object
(component E) near the center of the brightest cluster galaxy (G1)
confirms that it is the central fifth image of the lensed quasar
system. In addition, we measure the stellar velocity dispersion of G1
to be $\sigma_*=352\pm13$ km~s$^{-1}$. We combine these results to
obtain constraints on the mass $M_{\rm BH}$ of the putative black hole
(BH) at the center of the inactive galaxy G1, and hence on the $M_{\rm
  BH}$-$\sigma_*$ relation at the lens redshift $z_l=0.68$. From
detailed mass modeling, we place an upper limit on the black hole
mass,  $M_{\rm BH}<2.1\times10^{10}M_\odot$ at 1$\sigma$ level
($<3.1\times10^{10}M_\odot$ at 3$\sigma$),  which is consistent with
black hole masses expected from the local and redshift-evolved  $M_{\rm
  BH}$-$\sigma_*$ relations, $M_{\rm BH}\sim 10^9-10^{10}M_\odot$. 
\end{abstract}

\section{Introduction}

Quasars lensed by foreground clusters of galaxies serve as a powerful
probe of the mass distributions of clusters. SDSS J1004+4112 is the
first example of such quasar-cluster lens systems
\citep{inada03,oguri04,ota06}. It consists of four bright quasar
images ($z=1.734$) with a maximum image separation of $14\farcs6$  
produced by a massive cluster at $z=0.68$.  In addition to the quasar
images, multiply-imaged background galaxies are also observed
\citep{sharon05}. An advantage for using lensed quasars is that
they can provide time delays, which have been actually detected for
this system \citep{fohlmeister07,fohlmeister08}, containing unique
information on the lens potential.   

What makes the quasar lens SDSS~J1004+4112 particularly unique is the
probable central fifth image (component E) of the lensed quasar system.
High-resolution {\it Hubble Space Telescope (HST)} Advanced Camera for
Survey (ACS; \cite{clampin00}) images show a blue point source located
$0\farcs2$ from the center of the brightest cluster galaxy G1
\citep{inada05}. The central lensed image, if confirmed, has several
important implications for the central structure of lensing objects. 
For instance, central images can constrain the inner mass profiles of
lensing galaxies, particularly the masses of the central supermassive
black holes (e.g., \cite{mao01}; \cite{rusin01}; \cite{keeton03}; 
\cite{rusin05}).  While the
correlation between the black hole mass ($M_{\rm BH}$) and the stellar
velocity dispersions ($\sigma_*$) of the host galaxies has been
established for local galaxies
\citep{ferrarese00,gebhardt00,tremaine02}, measurements of the 
redshift evolution for the correlation, derived for galaxies hosting
active galactic nuclei, are still controversial (e.g., \cite{peng06}; 
\cite{shen08}; \cite{woo08}). Thus independent constraints from
central images are thought to provide insights into this relation.   

In this {\it Letter}, we present results of two deep spectroscopic
follow-up observations for SDSS J1004+4112, conducted at the Subaru
8.2-meter telescope. One is spectroscopy of component E to confirm 
its lensing nature, and the other is spectroscopy of galaxy G1 to 
measure its velocity dispersion. The latter observation is particularly 
important in separating the mass distribution of dark matter from that 
of baryons (e.g., \cite{sand08}). We then attempt mass modeling of the
lens system. Together with the measurement of $\sigma_*$ for galaxy
G1, it provides a direct constraint on the $M_{\rm BH}$-$\sigma_*$
relation at the lens redshift of $z=0.68$. In what follows, we assume
a standard flat universe with $\Omega_M=0.26$, $\Omega_\Lambda=0.74$,
and $H_0=72{\rm km\,s^{-1}Mpc^{-1}}$ (e.g., \cite{tegmark06}). 
 
\section{Observations and Data Analysis}

\begin{figure}
  \begin{center}
    \FigureFile(60mm,60mm){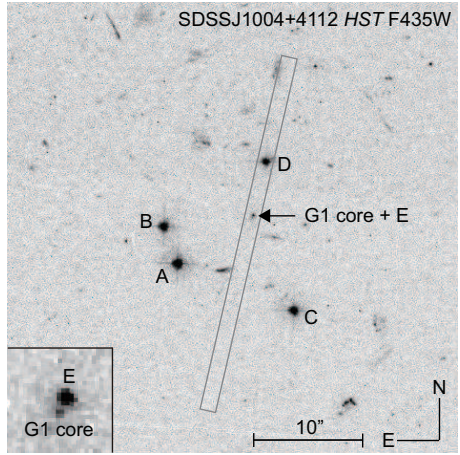}
  \end{center}
  \caption{
  {\it HST} ACS F435W image of SDSS~J1004+4112 with the exposure time
  of 5400~sec, taken on 2005 December 19 under the GO-10509 program
  (PI: C. S. Kochanek).  Components labeled by A--D indicate the 4
  lensed quasar images, and ``G1 core + E'' indicates the central
  fifth image E and the center of galaxy G1. The gray rectangle shows
  the slit direction for spectroscopy of the fifth image. The inset
  shows an expanded view of the central region of galaxy G1. North is
  up and East is left, and the pixel scale is 0\farcs05
  pixel$^{-1}$. See \citet{inada05} for {\it HST} images in other
  (redder) bands.   
  }
\label{fig:1004B}
\end{figure}

\subsection{Fifth Image}

\begin{figure}
  \begin{center}
    \FigureFile(60mm,60mm){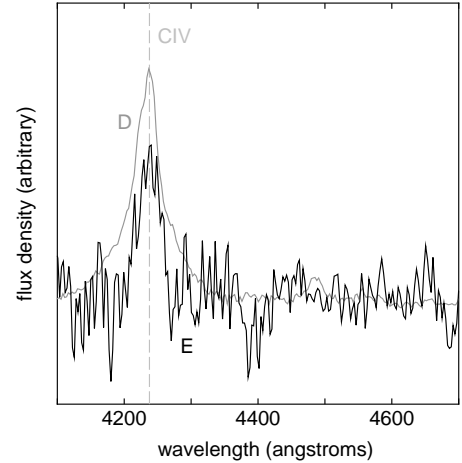}
  \end{center}
  \caption{
  Spectra of components E ({\it black}) and D ({\it gray}), normalized
  by each continuum level. In component E, an emission line is
  detected at the same wavelength as the C\emissiontype{IV} emission
  line of component D (also indicated by a vertical dotted line). The
  origin of the absorption features around  4200{\AA} and 4400{\AA} are
  unknown. 
  }
\label{fig:1d_DE}
\end{figure}

We conducted spectroscopic observations of the central fifth 
image (component E; see Figure \ref{fig:1004B}) with the Faint Object
Camera And Spectrograph (FOCAS; \cite{kashikawa02}) at the Subaru
8.2-meter telescope on 2007 January 22. We used the 300B grism, the
L600 filter, and a $1\farcs0$-width slit, under the 2$\times$2 
(spatial{$\times$}spectral) on-chip binning mode. With this 
configuration the wavelength coverage is from 3650{\AA} to 6000{\AA}, 
which covers a strong emission line (C\emissiontype{IV}) of the lensed
quasar with a spectral resolution of $\hbox{R}\sim400$ and a 
pixel scale of $0\farcs207$ pixel$^{-1}$ . 
Although component E is quite close to the center of galaxy G1
at $z=0.68$ ($0\farcs2$ from the center of G1), our blue spectroscopy
minimizes the contamination from G1. The slit direction was aligned to
pass through component D ($\sim 10\farcs0$ from component E, see Figure 
\ref{fig:1004B}) of SDSS J1004+4112 which we adopt as a reference for 
the quasar emission lines. Given the faintness ($B\sim24.5$) of 
component E, we used a total exposure time of 16,200~sec 
taken in excellent seeing conditions (FWHM of $\sim 0\farcs6$). After
removing cosmic rays using the Lacos\_spec task \citep{dokkum01}, we
extracted 1-dimensional spectra of components E and D using the
standard IRAF tasks. The spectra, normalized by the continuum level
of each image, are shown in Figure \ref{fig:1d_DE}. We find that the 
spectrum of component E shows a clear emission line just at the 
wavelength of the C\emissiontype{IV} emission line of component D. 
We note that there is no sky emission line near the wavelength of the
C\emissiontype{IV} emission line (4200{\AA}). Thus our spectroscopic
observations unambiguously confirm that component E is the central
fifth image of the lens system, representing the first spectroscopic
confirmation  of central odd images among lensed quasars. 

\subsection{Velocity Dispersion of G1}

To measure the velocity dispersion of galaxy G1, we also used FOCAS
but with a different setup. We adopted the 600-650 VPH grism 
(High 650), the SY47 filter, and a $0\farcs4$-width slit. The
wavelength coverage in this setup is from 5300{\AA} to 7700{\AA},
covering the Ca\emissiontype{II} H\&K and G-band absorption lines
of galaxy G1 at $z=0.68$. The slit direction was aligned along the
major axis of galaxy G1 (160.8 deg East of North). The spectrum, with 
a total exposure of 21,600~sec, was
obtained on 2007 January 21, under the 3$\times$1 on-chip binning
mode (a spatial scale of $0\farcs311$ pixel$^{-1}$). The setup was
chosen to achieve a spectral resolution of $\hbox{R}\sim2500$. 
The seeing was $\sim 0\farcs7$ FWHM. Again, cosmic rays were removed
by the Lacos\_spec task and the 1-dimensional spectrum of galaxy G1
was extracted by the standard IRAF tasks. 

We measure the velocity dispersion of G1 following the procedure
described in \citet{falco97}. We adopted HD~8491, HD~83085, HD~94247,
and HD~126778 in the coud\'{e} feed spectral library \citep{valdes04}
as template stars. We rebinned and smoothed the template spectra 
to match the observed spectral resolutions of our data. Both the
spectrum of G1 and the resolution-matched template stars are normalized
by the continuum levels. We then apply the Fourier cross-correlation
method \citep{tonry79} of the IRAF FXCOR task,
to the templates convolved with Gaussians. The resultant calibration
curves describe the relation between the input velocity dispersion and
the FWHM of the output cross-correlation peak measured  by the FXCOR
task. We used velocity dispersions from 200~${\rm km\,s^{-1}}$
to 450~${\rm km\,s^{-1}}$ with an interval of 10~${\rm
km\,s^{-1}}$. The FXCOR task was also applied to the normalized
rest-frame spectrum of G1 to obtain the width of the G-band absorption
line\footnote{The G-band is the only line which we can use, because the 
Ca\emissiontype{II} H\&K absorption lines are not suitable for
measuring velocity dispersions \citep{tonry98,ohyama02}.}, which
we convert to the velocity dispersion using linear interpolation
of the calibration curves. We find that 4 template stars  
yield similar velocity dispersions for G1 of 338--367~${\rm km\,s^{-1}}$.
From the average and scatter (no weight for the 4 independent measurements), 
we determine the velocity dispersion to be $\sigma_*=352\pm13$~${\rm
  km\,s^{-1}}$. Our measurement is consistent with $\sigma_*$
expected from the observed Faber-Jackson relation, $\sim320\pm 40$~${\rm
  km\,s^{-1}}$ \citep{liu08}.  

\section{Constraints on the $M_{\rm BH}$-$\sigma_*$ Relation}

The location and the brightness of the confirmed central image
allow us to place constraints on the mass of the supermassive BH
hosted by G1. This is made possible by the fact that central point
masses  (de-)magnify or even suppress central images (e.g.,
\cite{mao01}; \cite{rusin05}). Indeed the technique was applied to
three-image galaxy-scale lens PMN J1632$-$0033 to derive an upper
limit on $M_{\rm BH}$ in the lensing galaxy \citep{winn04}, although
the lack of the redshift measurement for the lensing 
galaxy makes the interpretation of this result somewhat difficult. 

Before conducting detailed mass modeling, we can estimate the BH mass 
range from the distance between component E and the center of G1. 
It is given by the mass at which the Einstein radius of the
central BH becomes comparable to this distance of $\sim 0\farcs2$,
corresponding to a BH mass of $M_{\rm BH}\sim 1.5\times10^{10}M_\odot$. 
Thus, we naively expect $M_{BH} \lesssim 10^{10} M_\odot$. 

To obtain more quantitative constraints, we fit the quasar images
with the following mass model. We adopt an elliptical version of the
(\cite{navarro97}; NFW) density profile for the dark matter distribution of 
the lensing cluster. Changing the inner slope of the dark matter
component is expected to have little effect on the fifth image,
because the center of the dark halo appears to be offset from G1
\citep{oguri04}.  The central position, ellipticity, position angle, 
concentration parameter, and total mass of the dark matter distribution 
are treated as free parameters. The central galaxy G1 is assumed to be 
an isothermal ellipsoid with varying core radius. We constrain the 
position angle to coincide with that of the light ($161\pm10$~deg 
East of North), but with no prior on the ellipticity. In addition we
added a Gaussian prior on the velocity dispersion of the isothermal
ellipsoid G1 from our measurement, $\sigma=352\pm13$~${\rm
km\,s^{-1}}$. The 16 galaxies other than G1, which we identify as
cluster members using their colors, are included as truncated
isothermal ellipsoids with an exact scaling relation, 
$\sigma_* \propto L^{1/4}$ and  $r_{\rm trunc}\propto L^{1/2}$. 
The overall normalization of $r_{\rm trunc}$  is left as a 
free parameter. We take the ellipticities and position angles for
these member galaxies from measurements in the {\it HST} ACS images.
Finally, we allow for an external shear following \citet{oguri04} to
achieve better fits. For observational constraints, we adopt the
positions and position errors of five quasar images and 
G1 derived in \citet{inada05}. We also include flux ratios of quasar
images as constraints, but with large errors of 
$\sigma(m_{\rm X}-m_{\rm A})=0.3$ (X represents images B--D) and
$\sigma(m_{\rm E}-m_{\rm A})=0.8$, in order to allow small flux
ratio changes due to time delays and microlensing. In addition, 
observed time delays, $\Delta t_{\rm AB}=40.6\pm1.8$~days 
\citep{fohlmeister07} and $\Delta t_{\rm
AC}=821.6\pm2.1$~days \citep{fohlmeister08}, are included. 

\begin{figure}
  \begin{center}
    \FigureFile(60mm,60mm){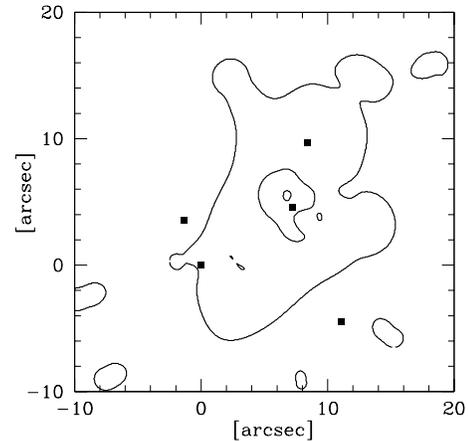}
  \end{center}
  \caption{The critical curves of our best-fit mass model. The
  observed positions of five quasar images are marked by filled
  squares, which are reproduced by our model nearly perfectly. 
  The best-fit parameters are $M_{\rm NFW}=2\times10^{15}M_\odot$,
  $c_{\rm NFW}=2$, $x_{\rm NFW}=6\farcs184$, $y_{\rm NFW}=6\farcs639$, 
  $e_{\rm NFW}=0.3$, $\theta_{\rm NFW}=-11^\circ$, $e_{\rm G1}=0.71$,
  $\theta_{\rm G1}=152^\circ$, $\sigma_{\rm G1}=355{\rm km\,s^{-1}}$, 
  $r_{\rm core,G1}=0\farcs1$, $\gamma=0.21$, $\theta_\gamma=82^\circ$, 
  and $r_{\rm trunc}(L=L_{\rm G1})=3\farcs2$.}
\label{fig:mbh_crit}
\end{figure}

\begin{figure}
  \begin{center}
    \FigureFile(70mm,70mm){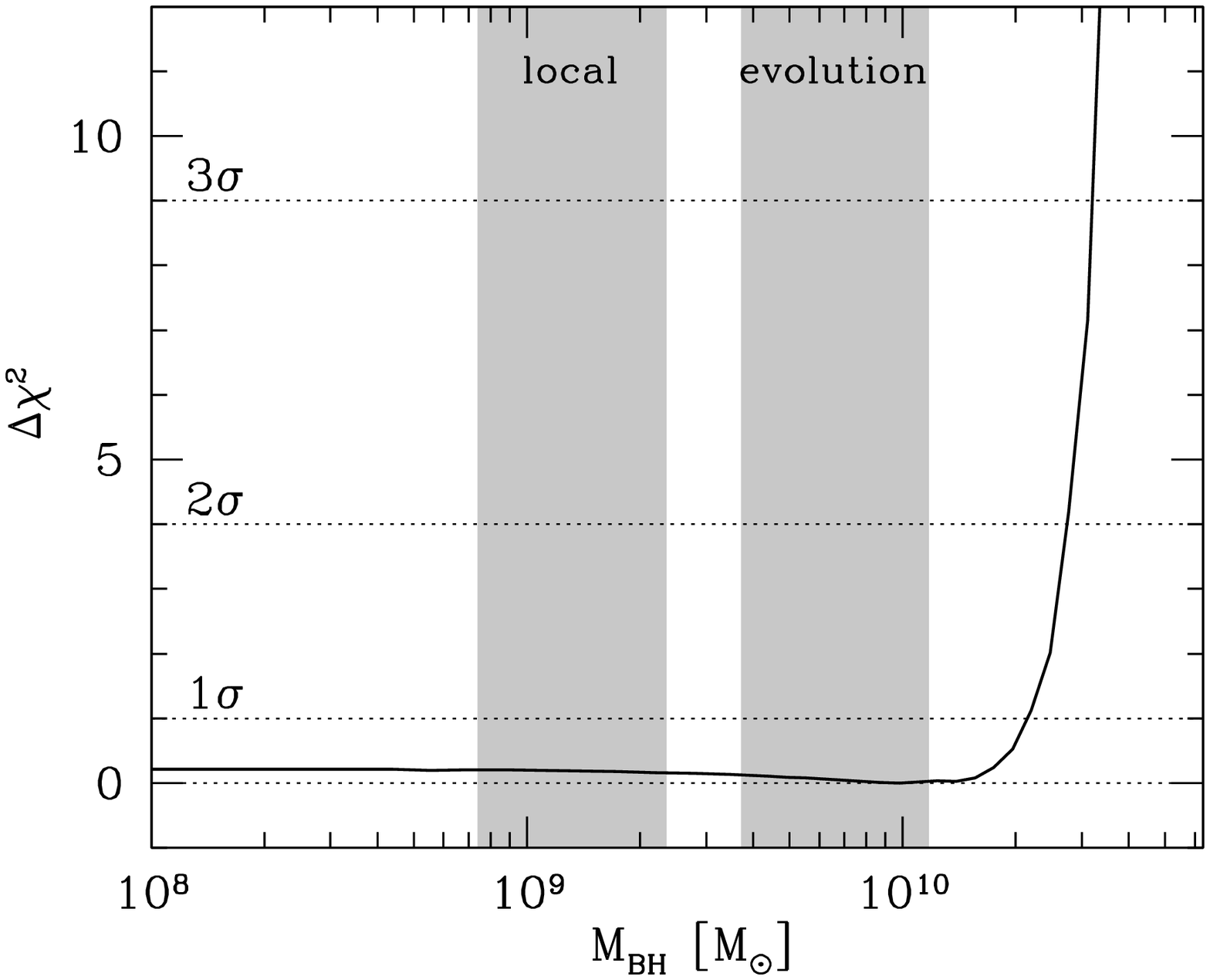}
  \end{center}
  \caption{The $\chi^2$ difference ($\Delta\chi^2$) as a function of the
  BH mass $M_{\rm BH}$. The steep rise of $\Delta \chi^2$ provides
  upper limits on $M_{\rm BH}$. The shaded regions indicate the values of
  $M_{\rm BH}$ expected from the measured stellar velocity dispersion
  $\sigma_*=352\pm13$ km s${}^{-1}$: The ``local'' value is estimated from
  the $M_{\rm BH}$-$\sigma_*$ relation of \citet{tremaine02}, 
  $M_{\rm BH}=10^{8.13}(\sigma_*/200{\rm km\,s^{-1}})^{4.02}M_\odot$ with an
  intrinsic dispersion of $\sim 0.25$~dex, whereas the ``evolution''
  value is computed in the same way as ``local'', except
  for an additional offset of the BH mass due to a possible redshift
  evolution measured by \citet{woo08}, $\Delta \log M_{\rm
  BH}=3.1\log(1+z)=0.70$ for $z=0.68$. }\label{fig:mbh}
\end{figure}

We find that our model reproduces the data quite well (see Figure
\ref{fig:mbh_crit} for the critical curves of our best-fit
model). With no BH, the best-fit chi-square is $\chi^2=2.57$ for 3
degree of freedom (20 constraints and 17 parameters). Thus the central
BH is not required to fit the fifth image. We then place
the BH (modeled by a point mass) at the center of galaxy G1, and we
re-perform $\chi^2$ minimizations  for each $M_{\rm BH}$. We note that
the central BH can produce a sixth image near the BH. In what follows
we ignore the sixth image as it is much fainter than the observed
fifth image in most of the situations considered here. 
We derive the upper   
limit of $M_{\rm BH}$ from the $\chi^2$ differences,
$\Delta\chi^2=\chi^2(M_{\rm BH})-\chi^2_{\rm min}$. Figure
\ref{fig:mbh} plots $\Delta\chi^2$ as a function of $M_{\rm  BH}$,
from which we derive constraints on the BH mass of $M_{\rm
  BH}<2.1\times10^{10}M_\odot$ at 1$\sigma$ and $M_{\rm
  BH}<3.1\times10^{10}M_\odot$ at 3$\sigma$.  The steep increase of
$\chi^2$ for large $M_{\rm BH}$ means that the mass of the BH is
limited to being a small fraction of the stellar mass interior to
image E.  

As shown in Figure \ref{fig:mbh}, our constraints on $M_{\rm BH}$ are
consistent with the expected local $M_{\rm BH}$-$\sigma_*$ relation
\citep{tremaine02} and also those inferred from the extrapolation of
a possible redshift evolution measured by \citet{woo08}. Although the
constraints are model-dependent, in the sense that they are derived
assuming a specific parametric mass model, we note that our mass model
is quite flexible near the center of G1 because the core radius is a
free parameter. Indeed, the best-fit core radius is correlated with
$M_{\rm BH}$; the core radius evolves from $\sim 0\farcs1$ for 
$M_{\rm  BH}=0$ to $\sim 0\farcs3$ for 
$M_{\rm   BH}=3\times 10^{10}M_\odot$. The rough agreement of this
constraint with a simple order-of-magnitude argument also suggests
that an upper limit to the BH mass of approximately $2\times
10^{10}M_\odot$ is generic. However, improved constraints may be
obtained by adding  constraints on the sixth image which we ignored,
because the sixth image becomes brighter with increasing $M_{\rm BH}$.
For instance, the flux of the sixth image is $\sim 20\%$ of E in our
best-fit models for $M_{\rm BH}\sim2\times 10^{10}M_\odot$, whereas
the {\it HST} image suggests that it cannot be $\gtrsim 10\%$ of E.   

\section{Summary}

We have presented the results of two spectroscopic observations at the
Subaru telescope. With the first observation, we confirmed that the
central point source found by \citet{inada05} is indeed the central 
fifth image of the lensed quasar by detecting the
C\emissiontype{IV} emission line of the quasar. This represents the first
spectroscopic confirmation of the central odd image. 
In the second spectroscopic observation, we determined
the stellar velocity dispersion of galaxy G1 to be
$\sigma_*=352\pm13$~${\rm km\,s^{-1}}$. 

We used these results to derive constraints on the $M_{\rm
  BH}$-$\sigma_*$ 
relation at the lens redshift $z=0.68$, assuming the presence
of a BH at the center of the brightest cluster galaxy G1. With a
parametric model which successfully reproduces the model constraints, 
we obtained limits of $M_{\rm  BH}<2.1\times10^{10}M_\odot$ at 1$\sigma$ and
$M_{\rm BH}<3.1\times10^{10}M_\odot$ at 3$\sigma$ that are consistent
with $M_{\rm BH}$ expected from the extrapolation of the known $M_{\rm
BH}$-$\sigma_*$ relation. It is worth noting that these constraints
are derived for an ``inactive'' galaxy with no nuclear activity. 
Current studies of the redshift evolution of the $M_{\rm
BH}$-$\sigma_*$ relation make use of the Balmer line widths, and therefore
are restricted to BHs in active galaxies. Given possible differences
in the $M_{\rm BH}$-$\sigma_*$ between active and inactive local
galaxies \citep{greene06}, constraints on $M_{\rm BH}$ for distant
inactive galaxies from gravitational lensing will be essential to
fully understand the origin of the relation.

\bigskip

N.~I. acknowledges support from the Special Postdoctoral Researcher 
Program of RIKEN. 
This work was in part supported by Department of Energy contract
DE-AC02-76SF00515. C.~S.~K. acknowledges support from NSF grant AST 
07-08082. G.~P.~S acknowledges support from a Royal Society University 
Research Fellowship. E.~F. acknowledges support from the Smithsonian 
Institution.

\end{document}